\documentclass[conference]{IEEEtran}
\usepackage{amsfonts}
\usepackage{bbm}
\usepackage[utf8]{inputenc}
\usepackage{epstopdf} 
\usepackage{epsfig}
\DeclareGraphicsExtensions{.eps}
\usepackage{tikz}
\usetikzlibrary{matrix,shapes,arrows,positioning,chains}

\usepackage{amsmath, amsthm, amssymb}

\usepackage{cite}
\usepackage{subfigure}
\usepackage{graphicx}
\usepackage[export]{adjustbox}

\usepackage{changes}

\ifCLASSINFOpdf
\else
\fi

\begin{document}

\title{Performance Analysis of Energy Consumption in Cache-Enabled Multicast D2D Communications}

\author{Bao~Wenhuan,\IEEEauthorrefmark{1}
        Mansour~Naslcheraghi,\IEEEauthorrefmark{2}
        and~Mohammad~Shikh-Bahaei,\IEEEauthorrefmark{1}
        \\
        \IEEEauthorblockA{Informatics center, King’s College London, London, UK \\
        	Email: \IEEEauthorrefmark{1}\{wenhuan.bao,m.sbahaei\}@kcl.ac.uk,
        \IEEEauthorrefmark{2}m.naslcheraghi@ieee.org}
}

\maketitle

\begin{abstract}
Device-to-Device (D2D) communication as a promising technology in 5G cellular networks provides the communication of the users in the vicinity and thereby decreases end-to-end delay and power consumption. In addition to the aforementioned advantages, it also supports the high-speed data transmission services such as content delivery. In this paper, we consider the D2D multicast communications opportunity in the D2D-cellular hybrid network, in which that one transmitter targets multiple receivers at the same time. We provide the analysis for the proposed system by using tools from stochastic geometry, to calculate the cache hitting probability of the receivers as well as the energy consumption of the hybrid network aiming to seek the optimal number of caching contents in the D2D multicast opportunities.
\end{abstract}


\IEEEpeerreviewmaketitle

\section{Introduction}\label{introduction}
In recent years, different multimedia services over mobile networks have attracted a great deal of attention and the demand for these servies are growing rapidly. To serve these tremendous demands, new intelligent user terminals such as smart phones and tablets provide users with new expreiences such as viewing video files with high quality. It can be observed that a vital part of the mobile video traffic is the repeated download of large size popular content (such as pop music or videos). In this regard, lots of contents are redundant and the traditional client-server model in the cellular network is not capable to serve these huge traffic. As a result, engineers and researchers adopt effective ways to reduce the duplicate video content delivery through using intelligent caching strategies. In this case, users can access video files pre-downloaded by their neighboring users using direct D2D communication. D2D multicast communication is becoming a support technology for video transmission in a wireless network to support users who need flexible QoS requirements. It allows serving multiple requests simultaneously by aggregating all these requests into one multicast stream \cite{CHEN20143}. To accomplish this, users should be encouraged in a way to be willing to contribute for sharing video files. That is to say, a mechanism should be introduced to convinced users to consume their batteries for sending video files to others. Network operators as one of the main participants and even stakeholders in serving traffic in cellular networks can provide these incentives for users and have their own benefit such as cellular traffic offloading \cite{article}. 

For a D2D wireless network, a suitable caching strategy is critical. It determines the probability that a user terminal can obtain the requested video files from the surrounding D2D user terminal cache, which is defined as cache hitting probability. When it comes to the existing cache strategies, Equal Probability Random Caching strategy (EPRC) and Most Popular File Caching (MPC) strategy are the most common ones \cite{7454446,ebenau2011genetic}. The first one means that all files are randomly cached by the D2D communication transmitters with the same probability. However, it does not take into account the diversity of user preference for popular files, which results in the low cache hitting probability. In the second one, all user terminals cache the same top-ranked file. When the file request is concentrated in the top of the file, a higher cache hitting probability could be achieved. Mansour NaslCheraghi concentrates on exploiting full-duplex (FD) D2D communication and proposes a new scheme for video distribution over cellular networks using tools from stochastic geometry. \cite{7998228,7933143,DBLP:journals/corr/NaslcheraghiGS17a}. Negin Golrezaei and Meiping Jia analyze the optimal collaboration distance in D2D unicast communication, trading off frequency reuse with the probability of finding a requested file within the collaboration distance \cite{6787081,9}. Yue Wang and Xiaofeng Tao consider the complex network scenario and use stochastic geometry for the closed form expression in order to improve successful transmission probability \cite{7843656}. Tao Zhang studies the caching cooperation scheme between hybrid-powered small base stations with the aim to minimize the system transmit power in heterogeneous networks \cite{7833620}. Rapeepat Ratasuk, Xingqin Lin and Amitabha Ghosh focus on multicast D2D communication. They proposed the tractable baseline multicast D2D and investigate some multicast factors such as mean number of covered receivers, throughput and coverage probability. Based on the characteristics of multicast and cooperation among BSs, Xiangyue Huang presents a caching structure and proposes a CMAC (Cooperative Multicast-Aware Caching) strategy to reduce the average latency of delivering content \cite{7881651}. On the other hand, abundant work exists on cross layer optimization of cellular networks through joint design of physical and MAC layers (e.g. see \cite{MSB1,MSB2,MSB3,MSB4,MSB5,MSB6,MSB7,MSB8}) . Some of these methods are applicable to our scenario as well. In this paper, we propose two caching strategies for D2D multicast communication. Details along with the main contributions are as follows:

\begin{itemize}
	\item First, we come up with the largest cache hitting probability in the hybrid cellular-D2D multicast network.
	\item Second, we analyze the overall throughput of the hybrid network to drive the ratio of energy consumption.
	\item Third, we compare these two models to obtain the optimal number of video files caching in the D2D multicast transmitters. 
\end{itemize}

The remainder of paper is structured as follows. In Section \ref{System Model} system model is introduced. In section III, the largest cache hitting probability model and the lowest energy consumption model are provided. In section IV simulation results are explained and conclusions are presented in section V.

\section{System Model}\label{System Model}
We consider a hybrid network in the cell consisting of both cellular and D2D multicast users. There is one base station, $\sigma$ cluster heads (the UTs cache the video content and prepare to transmit) and $\phi$ cluster members (the UTs need to download the video content) in the cell. The base station locates in the center. The positions of D2D multicast potential cluster heads form an independent Poisson Point Process distribution (PPP) $\Phi_{\sigma}=\sum_{i=1}^{\sigma}\sigma_i$ with intensity $\lambda_{\sigma}$; where $\sigma_i$ is binary 0-1 vector denotes the Dirac measure at position in the cell. Similarly, the positions of D2D multicast potential cluster members form an independent PPP $\Phi_{\phi}=\sum_{i=1}^{\phi}\phi_i$ with intensity $\lambda_{\phi}$. These two PPP distributions are independent of each other. Since a number of D2D receivers request same video files, the D2D multicast communication is suitable for popular video contents delivery. Even though discovery mechanism for the D2D communications don't affect our analysis, nevertheless, efficient D2D discovery mechanisms can be applied in system configurations \cite{MyDiscoveryD2D}. If the users are deterministic and the base station can obtain the location information of files caching, the best strategy is caching the most popular files in D2D transmitters evenly. Considering the Zipf distribution, each user terminal requests the video contents independently from the file library of video files $M=\{f_1, f_2,….f_m\}$. The probability of the video in higher ranking is also higher \cite{6495773,6195469}. When there are $M$ popular files in the file library, the probability of the file is requested in the $i^{\rm{th}}$ position can be expressed as:

\begin{align}
f_i = \frac{\frac{1}{i^{\gamma}}}{\sum_{j=1}^{M}\frac{1}{j^{\gamma}}}, i \in [1, M].
\end{align}


We assume that the cell is a circle with the radius of $R_C$. The base station is aware of the cached video files and channel state information of the user terminals and allocates sub-channels for D2D communication cluster, which adopts the centralized scheduling scheme. The number of total popular video files need to cache is $M$. Each potential D2D transmitter has a capacity of $\Omega$ files, and usually, $\Omega$ is no higher than $M (\Omega \leq M)$. Without loss of generality, we assume that $\frac{M}{\Omega}$ is an integer. In this case, $M$ video files need $\frac{M}{\Omega}$ D2D transmitters to cache them. In order to distribute the D2D transmitters in the cell relative evenly, we allow the repetition cache because the video files are very popular. Since we consider the large application area with the large density of D2D users, this assumption is feasible. Different video contents delivery is independent of each other. The next question is how to determine the optimal number of caching files $M_o$ in the cluster heads. Table \ref{mathematic_notations} gives a list of mathematic notations used in this paper

\begin{table}\label{mathematic_notations}
	\caption{List of notations}
	\resizebox{\columnwidth}{!}{
		\begin{tabular}{|c|l|}
			\hline
			Notation & Definition \\
			\hline \hline
			$R_C$ & Radius of the cell\\
			\hline
			$R_D$ & Radius of the D2D cluster \\
			\hline
			$M$ & The number of total video files \\
			\hline
			$\Omega$ & Cache capacity of each cluster head \\
			\hline
			$\sigma$ & The number of cluster heads \\
			\hline
			$\phi$ & The number of cluster members \\
			\hline
			$\lambda_\sigma$ & The intensity of cluster heads \\
			\hline
			$\lambda_\phi$ & The intensity of cluster members \\
			\hline
			$M_o$ & The optimal number of caching files \\
			\hline
			$P_h$ & The probability of file $i$ caching\\
			&  in the cluster heads \\
			\hline
			$P_{hit}$ & The hitting probability that cluster member \\
			& can find requested file $f_i$ \\
			\hline
			$p_i$ & The probability that the D2D cluster head \\
			& caching the file $i$ \\
			\hline
			$\sigma_a$ & The number of active cluster heads \\
			\hline
			$\phi_a$ & The number of active cluster members \\
			\hline
			$\mathcal{O}$ & The average capacity of each channel \\
			\hline
			$R_{EC}$ & The ratio of energy consumption \\
			\hline
			$\omega$ & The ratio of system EC  \\
			\hline
		\end{tabular} 
	}
\end{table}

\section{ANALYSIS}\label{Analysis}
\subsection{The Largest Hitting Probability (LHP)}\label{LHP}
We assume that there are $k$ D2D cluster heads within the D2D multicast range $R_D$ of one cluster member. According to the PPP formulation, the probability can be expressed as:
\begin{align}
P_r\{X=k\}=\frac{\lambda_\sigma^k}{k!}e^{-\lambda_\sigma}, \lambda_\sigma=\frac{\sigma \pi R_D^2}{\pi R_C^2}.
\end{align}
Since D2D user terminals are randomly distributed, cache hitting probability of each user terminal is the same. Therefore, we focus on one cluster member in the model. The probability of file $i$ caching in the cluster heads can be expressed as:
\begin{align}
\mathbb{E}[P_h|X=k]=1-(1-p_i)^k.
\end{align}
Where $p_i$ represents the probability that the D2D cluster head caching file $i$, so the number of files cache in the cluster head can not exceed the $\Omega$, i.e. $p_i=\frac{\Omega}{M},\sum_{i=1}^{M}p_i=\Omega$. Since there might be $\sigma$ cluster heads within the D2D communication range, according to (3-2), the average probability of file $i$ caching in the surrounding cluster heads can be expressed as:
\begin{align}
\mathbb{E}[P_h]=\sum_{k=0}^{\sigma}[1-(1-p_i)^k]P_r\{X=k\}.
\end{align}
The hitting probability that cluster member can find requested file $f_i$ in the cache is:
\begin{align}
\mathbb{E}[P_{hit}|f=f_i]=&\mathbb{E}[P_h]f_i=\sum_{k=0}^{\sigma}[1-(1-p_i)^k]P_r\{X=k\}f_i \notag &\\&
=\sum_{k=0}^{\sigma}\frac{\frac{1}{i^\gamma}}{\sum_{j=1}^{M}\frac{1}{j^\gamma}}\bigg[1-(1-\frac{\Omega}{M})^k\bigg]\frac{\lambda_\sigma^k}{k!}e^{-\lambda_\sigma}.
\end{align}

Since the size of the library is $M$, user terminal might request any video file in the library. We cache $M_o$ video files in the cluster heads. The average hitting probability that cluster member can find requested file $f_i$ in the cache is:
\begin{align}
\mathbb{E}[P_{hit}]&=\sum_{i=1}^{M_o}\sum_{k=0}^{\sigma}\frac{\frac{1}{i^\gamma}}{\sum_{j=1}^{M}\frac{1}{j^\gamma}}\bigg[1-(1-\frac{\Omega}{M})^k\bigg]\frac{\lambda_\sigma^k}{k!}e^{-\lambda_\sigma} \notag &\\&
=\sum_{i=1}^{M_o}\sum_{k=0}^{\sigma}\frac{\frac{1}{i^\gamma}}{\sum_{j=1}^{M}\frac{1}{j^\gamma}}\frac{\lambda_\sigma^k}{k!}e^{-\lambda_\sigma} \notag &\\& - \sum_{i=1}^{M_o}\sum_{k=0}^{\sigma}\frac{\frac{1}{i^\gamma}}{\sum_{j=1}^{M}\frac{1}{j^\gamma}}\bigg(1-\frac{\Omega}{M_o}\bigg) \frac{\lambda_\sigma^k}{k!}e^{-\lambda_\sigma} \notag &\\& = \sum_{i=1}^{M_o}\frac{\frac{1}{i^\gamma}}{\sum_{j=1}^{M}\frac{1}{j^\gamma}}e^{\lambda_\sigma}e^{-\lambda_\sigma}
-\sum_{i=1}^{M_o}\frac{\frac{1}{i^\gamma}}{\sum_{j=1}^{M}\frac{1}{j^\gamma}}e^{\lambda_\sigma(1-\frac{\Omega}{M})}e^{-\lambda_\sigma} \notag & \\ & =
\sum_{i=1}^{M_o}\frac{\frac{1}{i^\gamma}}{\sum_{j=1}^{M}\frac{1}{j^\gamma}} \bigg(1-e^{-\frac{\Omega}{M_o}\lambda_\sigma}\bigg).
\end{align}

Then we need to determine the optimal value of $M_o$ as we discussed before. In order to maximize the cache hitting probability $P_{hit}$, we should make full use of each cluster head. We assume that the cluster heads in a unit cell can cache the video popular with repetition. That is to say, cluster heads can cache repeated video files when $\sigma\Omega > M_o$, which can be expressed as:
\begin{align}
M_o & = \underset{M_o}{\mathrm{argmax}} \text{ } \mathbb{E}[P_{hit}] \notag & \\ &
= \underset{M_o}{\mathrm{argmax }} \sum_{i=1}^{M_o}\sum_{k=0}^{\sigma}\frac{\frac{1}{i^\gamma}}{\sum_{j=1}^{M}\frac{1}{j^\gamma}}\bigg[1-(1-\frac{\Omega}{M})^k\bigg]\frac{\lambda_\sigma^k}{k!}e^{-\lambda_\sigma} \notag &\\&
\text{Subject to: } \lambda_\sigma = \frac{\sigma \pi R_D^2}{\pi R_C^2}, M_o \in [\Omega, M].
\end{align}

\subsection{The Lowest Energy Consumption (LEC)}\label{LEC}
We determine that the probability of UT transmitting in D2D mode is $p_{D2D}$; the probability of UT transmitting in cellular mode is $p_C$. Since we permit the repetition of video file cache in the unit cell, the average probability of the cluster heads caching the needed files can be expressed as $P_i=\frac{\Omega}{M_o}$. The probability that there are cluster members requesting the video files is: 
\begin{align}
\mathbb{E}[P_{D2D}|f=f_i]=P_i.f_i.
\end{align}
The average probability that there are cluster members requesting the video files considering $M_o$ files cached in transmitters is:
\begin{align}
\mathbb{E}[P_{D2D}] = \sum_{i=1}^{M_o}P_i \frac{\frac{1}{i^\gamma}}{\sum_{j=1}^{M}\frac{1}{j^\gamma}}.
\end{align}

Hence, the average number of active clusters can be expressed as: 
\begin{align}
\mathbb{E}[\sigma = \sigma_a] & = \sigma \sum_{k=0}^{\phi}\bigg[1-(1-\mathbb{E}[P_D2D])^k\bigg]P_r\{Y=k\} \notag &\\& = \sigma \sum_{k=0}^{\phi} \Bigg[1-\bigg(1-\frac{\Omega}{M_o}\sum_{i=1}^{M_o}\frac{\frac{1}{i^\gamma}}{\sum_{j=1}^{M}\frac{1}{j^\gamma}}\bigg)\Bigg]\frac{\lambda_\phi^k}{k!}e^{-\lambda_\phi}.
\end{align}

Where $\lambda_\phi = \frac{\phi \pi R_D^2}{\pi R_C^2}$. Then we define that EC is the average energy consumption of user terminals, which can be expressed as the power and throughput i.e. $EC=\frac{P}{\mathcal{O}}$. We assume that the user terminal and the base station adopt a fixed transmit power. Since the user terminals are randomly distributed, the average transmission rate and energy consumption is the same. Therefore, we can focus on one user terminal. In the hybrid network, the average energy consumption is given by:

\begin{align}
EC_d = \mathbb{E}[P_{hit}]\frac{P_D}{\mathcal{O}_D} + (1 - \mathbb{E}[P_{hit}])\frac{P_C}{\mathcal{O}_C}.
\end{align}

Where $\mathcal{O}_D = \frac{\phi_a}{\sigma_a}\mathcal{O}_d$, $\mathcal{O}_d$ represents the average capacity of each channel of the D2D cluster, $\mathcal{O}_D$ and $\mathcal{O}_C$ represent the average throughput of the D2D network and cellular network, respectively. In the cellular network, the average energy consumption is:
\begin{align}
EC_c = \frac{P_c}{\mathcal{O}_c}.
\end{align}

The ratio of energy consumption is: 
\begin{align}
R_{EC} & = \frac{EC_d}{EC_c} = \sum_{i=1}^{M_o} \bigg(p_h f_i \omega \frac{\sigma_a}{\phi p_{hit}} + p_c f_i\bigg) + \sum_{i=M_o}^{M} p_c f_i \omega \notag & \\ & 
= \frac{P_D}{\mathcal{O}_d} / \frac{P_c}{\mathcal{O}_c} \notag & \\ & 
R_{EC} = 1 - \sum_{i=1}^{M_o}\sum_{k=0}^{\sigma} [1 - (1 - p_i)^k] P_r\{X=k\}f_i \notag & \\ & 
+ \frac{\sigma \omega}{\phi} \sum_{k=0}^{\phi} \Bigg[1 - \Big(1 - P_i \sum_{i=1}^{M_o}\frac{\frac{1}{i^\gamma}}{\sum_{j=1}^{M}\frac{1}{j^\gamma}}\Big)\Bigg] \frac{\lambda_\phi^k}{k!}e^{-\lambda_\phi} \notag & \\ & 
= 1 - \sum_{i=1}^{M_o}\frac{\frac{1}{i^\gamma}}{\sum_{j=1}^{M}\frac{1}{j^\gamma}} \bigg(1-e^{-\frac{\Omega}{M_o}\lambda_\sigma}\bigg) \notag & \\ & 
+ \frac{\sigma \omega}{\phi} \bigg(1-e^{-\frac{\omega}{M_o}\sum_{i=1}^{M_o}\frac{\frac{1}{i^\gamma}}{\sum_{j=1}^{M}\frac{1}{j^\gamma}}} \bigg).
\end{align}

In this case, the optimal number of video cache files can be written as:

\begin{align}
M_o & = \underset{M_o}{\mathrm{argmax}} \text{ } R_{EC} \notag & \\ &
\text{Subject to: } \omega = \frac{P_D}{\mathcal{O}_d} / \frac{P_c}{\mathcal{O}_c}, \lambda_\sigma = \frac{\sigma \pi R_D^2}{\pi R_c^2}, M_o \in [\Omega, M].
\end{align}

\section{Numerical Results}\label{Results}
In this section, we present simulation results evaluating the performance of our two proposed cache strategies. Simulation parameters are shown in Table \ref{Parameters_table}.

\begin{table}
	\caption{Simulation Parameters}
	\label{Parameters_table}
	\resizebox{\columnwidth}{!}{
		\begin{tabular}{|c|c|c|}
			\hline
			Symbol & Parameter & Value\\
			\hline \hline
			$R_C$ & \text{Radius of the cell} & \text{200m} \\
			\hline
			$R_D$ & \text{Radius of the D2D cluster} & \text{50m} \\
			\hline
			$M$ & The number of total video files & \text{500}\\
			\hline
			$\Omega$ & \text{Cache capacity of each cluster head} & \text{10} \\
			\hline
			$\sigma$ & \text{The number of cluster heads} & \text{100} \\
			\hline
			$\omega$ & \text{The ratio of system EC factor} & \text{0.1} \\
			\hline
			$\phi$ & \text{The number of cluster members} & \text{250} \\
			\hline
		\end{tabular}
	}{\vspace{-5mm}}
\end{table}

\begin{figure}
	\centering
	\resizebox{\columnwidth}{!}{
		\includegraphics[width=0.6 \textwidth]{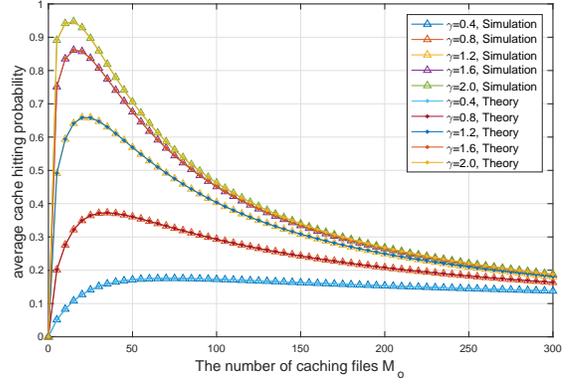}
	}
	\caption{The average cache hitting probability versus the number of caching files}
	\label{2}
\end{figure}

As shown in Fig. \ref{2}, with the increase of $\gamma$, the maximum average cache hitting probability is increasing at the same time. Because D2D cluster members request the more popular video files thereby, the probability of requesting the video files with high ranking is increasing. Therefore, we only need to cache these popular files to realize the higher cache hitting probability. Caching one tenth of total video files can obtain the highest probability when $\gamma$ is more than 1. On the other hand, this cache strategy relies on the value of the exponential constant of the Zipf distribution. The small value of $\gamma$ is corresponding to the low average cache hitting probability. 

\begin{figure}
	\centering
	\resizebox{\columnwidth}{!}{
		\includegraphics[width=0.6 \textwidth]{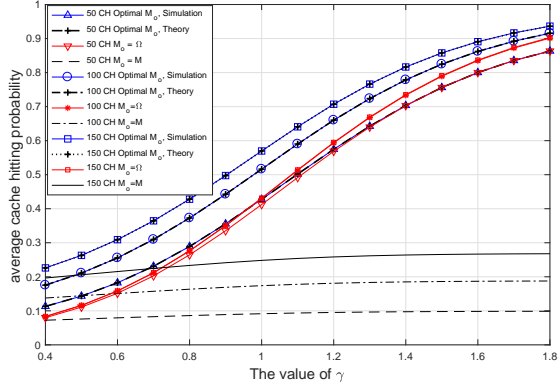}
	}
	\caption{The average cache hitting probability versus the value of $\gamma$ for different cache strategies and number of cluster heads}
	\label{3}
\end{figure}

The average cache hitting probability (CHP) versus $\gamma$ is shown in Fig. \ref{3}. We can observe that it increases when $\gamma$ increases as we discussed before. On the other hand, more cluster heads usually mean the larger hitting probability because there are more cluster heads which can be potential D2D transmitters around the cluster member requesting the video files. When we cache all of the video files (the maximum value of the library) in the D2D cluster heads, the CHP tends to be very small and smooth. Because a large amount of unneeded video files waste the limited cache ability. We can notice that two curves coincide when we only cache the most popular video files which equal to cache capacity of the cluster heads. Hence, the number of the cluster heads meet the basic need of cluster member, that is to say, there must be at least one cluster head around the cluster member within the D2D communication radius on average. The blue curves show that adopting LHC strategy can obtain the highest CHP.

\begin{figure}
	\centering
	\resizebox{\columnwidth}{!}{
		\includegraphics[width=0.6 \textwidth]{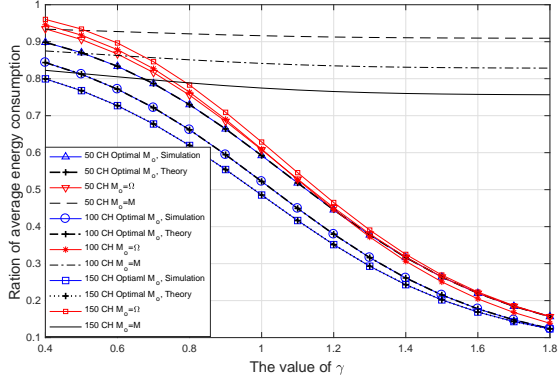}
	}
	\caption{The average ratio of energy consumption (REC) versus the value of $\gamma$ for different cache strategies and number of cluster heads}
	\label{4}
\end{figure}

Fig. \ref{4} shows the average ratio of energy consumption (REC) versus $\gamma$. Contrary to the previous picture, we can observe that it decreases as $\gamma_r$ increases. In the case of $(\gamma,CH)=(1,100)$ and $(\gamma,CH)=(1,150)$, the system saves energy consumption about $48\%$ and $52\%$, respectively. When $\gamma=1.4$, the system can save energy consumption more than $70\%$. We can conclude that D2D multicast communication has large potential in the aspects of saving network system energy. Different from the Fig. \ref{2}, two curves do not coincide when we only cache the most popular video files which equal to cache capacity of the cluster heads. Because we need relatively smaller amount of active cluster heads in order to realize larger REC when the request of video files has been met.

\begin{figure}
	\centering
	\resizebox{\columnwidth}{!}{
		\includegraphics[width=0.6 \textwidth]{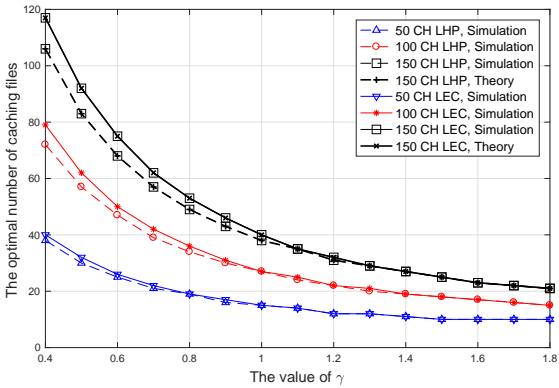}
	}
	\caption{The number of optimal caching files versus the value of $\gamma$ for different cache strategies and number of cluster heads}
	\label{5}
\end{figure}

Fig. \ref{5} shows the optimal number of caching video files. The value of $M_o$ decreases with the increase of $\gamma$. Since D2D cluster heads only need to cache the most popular video files to ensure the high CPH or the low REC. There is no doubt that more cluster heads can cache more video files, which can be effective to improve the CPH or reduce the REC when the value of $\gamma$ is relatively small. On the other hand, The LHP strategy caches a lower number of optimal caching files because it aims to obtain largest cache hitting probability and fewer video files can add the CHP. The LEC strategy tends to obtain larger ratio of energy consumption, that is to say, more video files caching gets more active D2D cluster heads. High value of $\gamma$ corresponds to the large number of the active clusters, because residents have a more centralized demand for video files. In addition, we can observe that the optimal number of caching files tends to be equal to the capacity of the cluster head. Hence, we will discuss about the scenarios that there are a number of cluster heads transmitting same video contents in next chapter.

\begin{figure}
	\centering
	\resizebox{\columnwidth}{!}{
		\includegraphics[width=0.6 \textwidth]{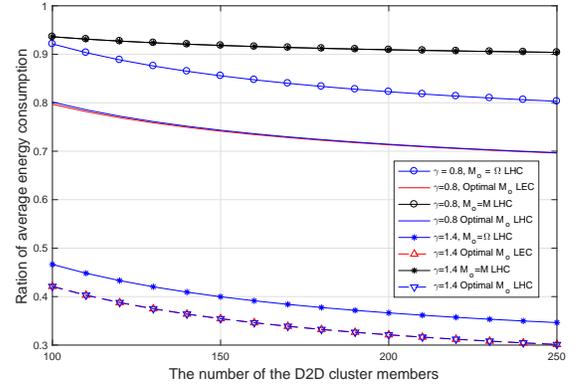}
	}
	\caption{The average ratio of energy consumption (REC) versus the number of the D2D cluster members for different cache strategies and value of $\gamma$}
	\label{6}
\end{figure}

Fig. \ref{6} shows the effect of different number of cluster members on REC. We can observe that the REC is almost not affected when we cache all of the video files in the D2D cluster heads. Caching optimal number of popular video files obtains the lowest REC compared to other two ways of caching strategy. The LEC strategy shows better performance than the LHC because LHC aims to obtain larger cache hitting probability.

\section{Conclusion}\label{Conclusion}
We propose the largest cache hitting probability model and the lowest energy consumption model to determine the optimal number of caching files in cluster heads. There is no doubt that D2D multicast communication is more suitable for the most popular video content delivery. The high value of $\gamma$ corresponds to the large number of the active clusters, because residents have a more centralized demand for video files. And the optimal number of caching files tends to be equal to the capacity of the cluster head when $\gamma$ is large enough. The future work should give priority to study the case with moving user terminals. The model we proposed in this paper focuses on the deterministic user terminals, which means that the base station needs to adopt these algorithms during each determined time slot that cause a large overhead of network control.

\tikzstyle{decision} = [diamond, draw,  
text width=5em, text badly centered, node distance=3cm, inner sep=0pt]
\tikzstyle{block} = [rectangle, draw,  
text width=6em, text centered, rounded corners, minimum height=4em]
\tikzstyle{line} = [draw, -latex']
\tikzstyle{cloud} = [draw, ellipse, node distance=3cm,
minimum height=2em]

\appendices
\bibliographystyle{IEEEtran}

\end{document}